\documentclass[aps,preprint,tighten,floats,epsf,rotate,showpacs]{revtex4}
\usepackage{graphicx}

\begin{document}
\title{Effect of citation patterns on network structure}
\author{Soma Sanyal}
\affiliation{School of Library and Information Science, 1320 East 10th Street, Bloomington
47405}

\begin{abstract}

We propose a model for an evolving citation network that incorporates the citation pattern followed 
in a particular discipline. We define the citation pattern in a discipline by three factors. 
The average number of references per article, the probability of citing an article based on it's age 
and the number of citations it already 
has. We also consider the average number of articles published per year in the discipline. 
We propose that the probability of citing an article based on it's age can be modeled by a 
{\it lifetime distribution}. The lifetime distribution models the citation lifetime of an 
average article in a particular discipline.   We find that the citation lifetime 
distribution in a particular discipline predicts the topological structure of the citation network 
in that discipline. We show that the power law exponent depends  
on the three factors that define the citation pattern. Finally we fit the data from the Physical 
Review D journal to obtain the citation pattern and calculate the total 
degree distribution for the citation network.


\end{abstract}

\maketitle

\section{Introduction}
\label{sec:I}

Citation networks of scholarly publications have generated considerable interest 
in recent times. In citation networks, the articles are the nodes and an edge is attached 
between two nodes when one article cites another. Similar to other networks, such as the 
telecommunication network, social networks, neural networks etc., citation networks were also observed 
to have a scale free structure \cite{price}. The first study of citation statistics was made by de Solla 
Price\cite{price} 
who proposed that the rate at which an article gets cited is proportional to the number of 
citations it already has. His model was based on the models by Yule \cite{yule} and Simon \cite{simon}.   
Later, a simple analytical model which leads to a scale free 
structure in growing networks was proposed by Barab\'{a}si and Albert \cite{barabasi1}.  
The Barab\'{a}si- Albert (BA) model introduced a similar concept called \textit{preferential attachment} whereby a new node
is connected to some old node in the network based on it's number of edges. This leads to the scale free 
structure in most networks. In all these models, an older node with a large number of edges will go on 
accumulating new edges irrespective of it's age.  However, 
citation networks obtained from real world datasets do not follow a simple scaling solution.
Though people tend to cite highly cited papers, old papers are rarely cited, especially 
if the research described in them have already been incorporated in recent text books. This 
lead to modifications to the BA model where the attachment probability depended on the 
number of edges of the node and was also proportional to some power of the age of the node 
\cite{dorogovtsev1}. 

Lehmann et. al. \cite{lehmann} found that though the citation 
network in the SPIRES database could be described by simple power laws, the exponent of the power law 
changed when the number of citations exceeded a certain value. The data thus suggested that the citation 
distribution is described by two independent power laws in two different domains. The data could 
also be fitted with a stretched exponential. They developed a model \cite{lehmann2} where some nodes are 
considered dead after sometime and are not able to acquire new links in the growing network.
Their model successfully explained the results from the SPIRES database.
Similarly Chun et. al. \cite{chun} who studied the 
citation network in the journal Scientometrics, found that the out-degree in the citation network 
could be fitted by two independent power laws whereas the in-degree could be fitted with a single power 
law. Vazquez \cite{vazquez} did a detailed study of the out-degree distribution of citation networks from 
various journals. He found that there was a difference in the citation distribution for journals which 
have a restriction in the number of pages per article. He proposed a recursive search model which partially 
explained some of the features observed in various citation networks.
B\"{o}rner et al.\cite{borner} proposed the TARL (Topics, 
Aging, Recursive Linking) model, which is a general process model that models the growth of a bipartite 
network of authors and articles. They used a 20 year data set of articles published in the Proceedings of 
the National Academy of Sciences {\it(PNAS)} to validate their model. The deviations from power law in the 
citation distribution observed for the PNAS data set could  be explained by their model. In the TARL
model, the probability of highly cited papers garnering more citations is offset by incorporating a bias 
to cite more recent papers. The number of citations received as a function of age was fitted by a Weibull 
distribution. The scale parameter of the Weibull distribution represented the aging bias. 

Apart from these models that explained the scaling co-effecients of specific data sets, there were 
other general analytical models where the attachment probability depended exponentially 
on the age of the node \cite{sen,zhu}. These studies demonstrated that the introduction of aging 
significantly transformed the statistical properties of the growing network \cite{zhu}. In this work, 
we aim to understand the underlying reasons that govern the power law behavior and scaling co-efficients 
of various citation networks.   We propose a general analytical model that incorporates the citation 
pattern in a particular discipline to predict the topological structure of the final network. 
Citation patterns in different disciplines are not always well defined. To obtain  quantitative estimates
we define the citation pattern in a particular discipline by the following factors, 
(i) the average number of references per article,
(ii) the citation probability based on age and (iii) the citation probability based on the  number of 
citations it already has. In addition, we also consider the average number of articles published 
in that discipline per year. Our citation probability depends on age and previous number of citations.
These two probabilities are assumed to be independent of each other. 
We use the master equation approach and solve the difference-differential equation to 
obtain the probability of a node to have $k$ number of citations at time $t$. Finally, the degree 
distribution is obtained by summing over all the nodes in the network at a particular time.

The probability of citing an article based on the number of citations it already has, is modeled 
by the preferential attachment method proposed by Barab\'{a}si and Albert \cite{barabasi1}. We propose that the 
probability of citing an article based on it's age can be modeled by a lifetime distribution. Lifetime 
distributions like the Weibull and the lognormal have been used extensively for life data analysis in 
predicting the mean life for a wide array of complex machinery and other products for everyday use. 
They have been used to model the lifetime of organisms in an ecosystem, here we
use them to model the citation lifetime of an article in a particular discipline.
Citation lifetime is given by the distribution of citations over time \cite{beaver}. A high citation 
lifetime indicates that articles are being cited over a longer period of time.  
The other reason for using lifetime distributions is that 
they are  very versatile and can take on the characteristics of other types of distributions 
based on the value of their parameters (e.g. the Weibull gives the exponential distribution for a 
shape parameter of $1$).      
We find that the probability of a node to have $k$ number of citations at time $t$ depends on the parameters 
of the chosen distribution. For certain parameters it is possible to obtain a power law distribution while 
for others it is more like a stretched exponential. Since the parameters of the chosen distribution themselves
depend on the citation pattern in that particular discipline, we conclude that it is the citation
pattern in a particular discipline which  plays a significant role in the degree distribution of a citation 
network. Our results show that the scaling exponents of the final degree distribution depends significantly on 
the citation pattern.

\section{Our Model}
\label{sec:II}

As we are dealing with a citation network here, the nodes of the network are the articles that are 
published. So nodes and articles refer to the same object in our model. 
Like previous models \cite{sen,zhu} we consider preferential attachment and aging 
to be independent of each other. So the total attachment probability in our 
model depends on the degree of a particular node $k$ and it's age $(t-t_0)$ where $t$ is the present 
time and $t_0$ is the time it entered the growing network.
We assign an initial degree $k_0$ to each node. $k$ is the total degree of a particular node, so its out-degree
is $k_0$ and in-degree is $(k-k_0)$. For the time being however, we do not make a distinction between 
incoming and outgoing degree distribution. Our time unit here is in years and $m$ number of nodes enter the 
network at each timestep. The probability $p(k,t-t_0)$ that a node (which had entered the 
network at time $t_0$) will have a degree $k$ at time $t$ is given by,   
\begin{equation}
  \label{eq:prob}
  p(k,t-t_0) = f(t-t_0)\left[\frac{k-1}{2t} p(k-1,t-t_0-1) - (1 - \frac{k}{2t})p(k,t-t_0-1)\right]
\end{equation}
where $f(t-t_0)$ is a lifetime distribution which satisfies the condition $f(t-t_0) = 0$ at $t = t_0$.

The master equation obtained from Eq. \ref{eq:prob} is given by,
\begin{equation}
  \label{eq:master}
  \left[\frac{2 t}{f(t-t_0)}\right][p(k,t-t_0+1)-p(k,t-t_0)] = (k-1) p(k-1,t-t_0) - k p(k, t-t_0). 
\end{equation}
We do a standard Z-transform,
\begin{equation}
  \label{eq:zdef}
   \Phi(z,t-t_0) = \sum p(k,t-t_0) z^{-k}
\end{equation}
and obtain, 
\begin{equation}
  \label{eq:ztrans}
  \left[\frac{2 t}{f(t-t_0)}\right]\frac{d\Phi}{dt} = (z-1)\frac{d\Phi}{dz}
\end{equation}
This is subject to the condition at $t = t_0$, 
\begin{equation}
  \label{eq:initcond}
  \Phi(z,0) = m z^{-k_0}, 
\end{equation}
where $m$ is the average number of articles published in that discipline per year and $k_0$ is the 
average number of references an article has. The other condition is,  
\begin{equation}
  \label{eq:bouncond}
 \Phi(1,t-t_0) = 1. 
\end{equation}
Eq. \ref{eq:ztrans} is solved using the above conditions and the final solution is given by, 
\begin{equation}
  \label{eq:soln}
p(k,t-t_0) = {\frac{(k_0 + k -1)!}{(k_0-1)!k!}}{(-1)^k} m\left[(\exp(-g(t-t_0))-1)^k \right]\left[\exp(-k_0 g(t-t_0))\right]
\end{equation}
where $g(t-t_0) = \int \frac{f(t-t_0)dt}{2t}$. If $f(t-t_0)$ is the Weibull distribution,
it is possible to solve this integral analytically 
using special functions. The final answer is a non-trivial combination of the error function and the 
exponential integral. Since we are dealing with finite time periods, we choose to solve the integral
numerically. 
 
In the next section, we obtain the total degree distribution for two different distributions, the 
Weibull distribution and the lognormal distribution.
 The two-parameter Weibull and the lognormal distribution are characterized by the 
shape parameter $a$, and the scale parameter $b$.  We impose the condition $f(t-t_0)$ is zero at 
$t = t_0$.  For a given dataset, we can fit the data 
and obtain the two parameters but for our analysis we consider the shape parameters to be constant and 
vary only the scale parameters. The effect of the scale parameter is to stretch the distribution. Since 
the distribution is over age here, increasing the scale parameter means citing relatively older papers.

\section{Results}
\label{sec:III}

Apart from the parameters characterizing the probability of citing articles based on their age, we also 
have two other inputs, the average number of papers published each year $m$, and the average number of 
references $k_0$  each paper has. Initially we keep these two factors constant and investigate the 
effect of the citation distribution on the total degree distribution. We consider $m = 200$ and 
$k_0 = 20$ and the time period is 50 years.

\subsection{Weibull distribution }
\label{sec:1}

We consider the two parameter Weibull distribution given by,
\begin{equation}
  \label{eq:weibull}
  f(t-t_0) = a b^{-a}{(t-t_0)}^{(a-1)} \exp(-{((t-t_0)/ b)^{a}})
\end{equation} 
The shape parameter $a = 2$ satisfies our initial condition. 
The scale parameter is varied to study how the distribution pattern affects the final structure of the 
network. 

\begin{figure}
\begin{center}
\includegraphics[angle=270,width=80mm]{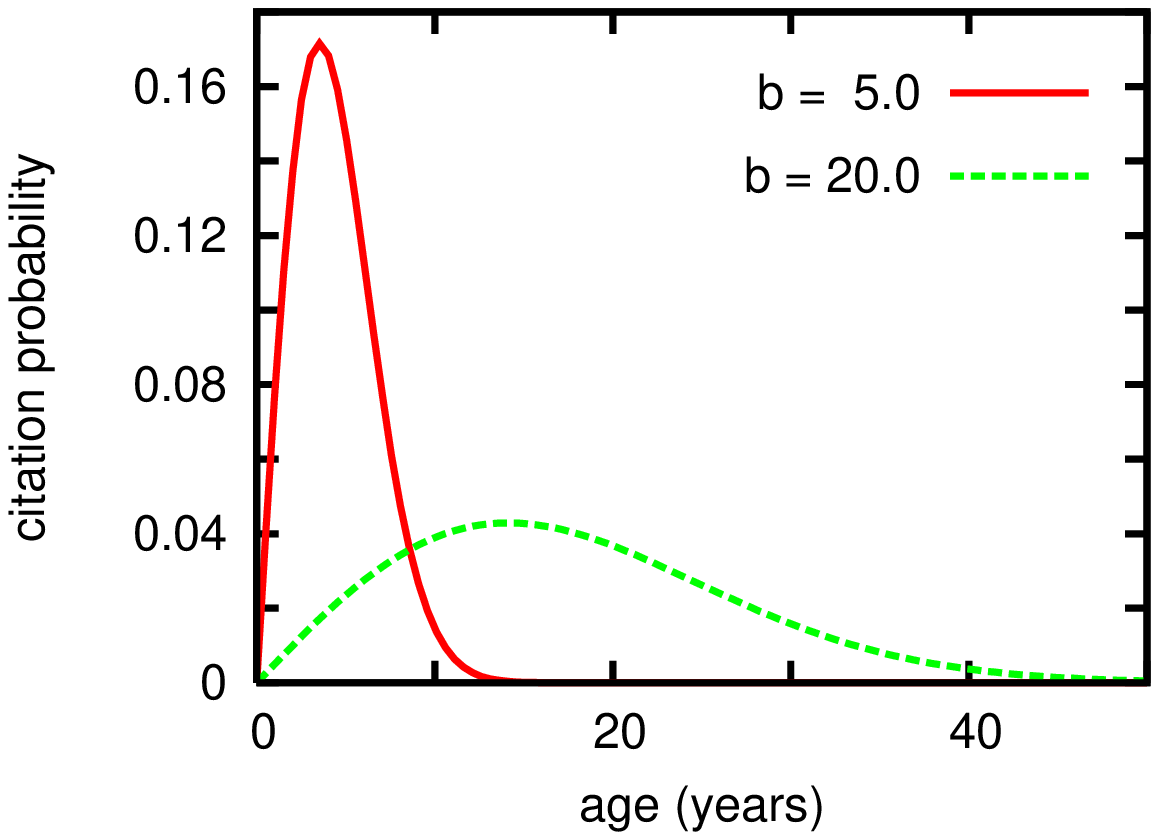}
\end{center}
\vskip -0.5cm
\caption{}{Citation lifetime distribution of an article for the Weibull distribution. 
The solid (red) line is for $b=5$ and the dashed (green) line is for $b=20$.}
\label{Fig.1}
\end{figure}
Fig. \ref{Fig.1} gives the citation lifetime distribution of an article for the Weibull distribution.
The x-axis is the age of the article and the y-axis gives it's probability of citation. 
The solid (red) line is for scale parameter $b=5$ and the dashed (green) line is for $b=20$.
$b=20$ means that the discipline tends to cite relatively older articles while $b = 5$ indicates that 
recent articles are more likely to be cited.

\begin{figure}
\begin{center}
\includegraphics[angle=270,width=80mm]{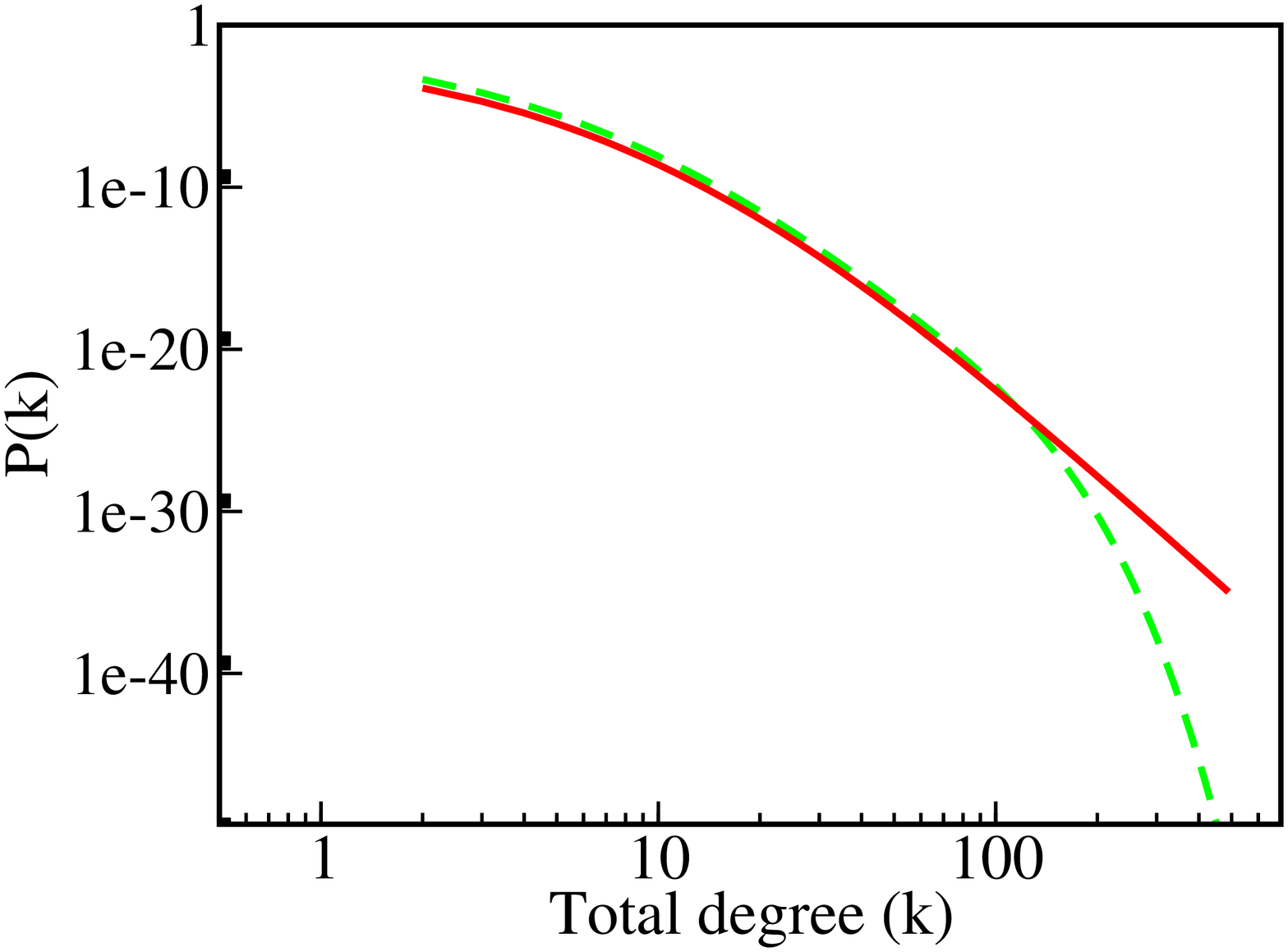}
\end{center}
\vskip -0.5cm
\caption{}{Degree distribution of citations to an article for different values of the scale parameter
in the Weibull distribution. The solid (red) line is for $b=5$ and the dashed (green) line 
is for $b=20$.}
\label{Fig.2}
\end{figure}

We obtain the total degree distribution for the two scale parameters in Fig. \ref{Fig.2}
For the Weibull function the tail of the degree distribution changes when the citation pattern changes.  
We do not get a simple power law in any of the cases.
However, for $b = 5$, it is possible to fit the final degree distribution using 
two power laws. This implies that in disciplines where more recent articles are likely 
to be cited, the total degree distribution can be described by two power laws with different 
coefficients.

\subsection{Lognormal distribution}
\label{sec:2}

The other commonly used lifetime distribution is the lognormal distribution given by, 
\begin{equation}
  \label{eq:lognormal}
  f(t-t_0) = \frac{\exp((-\ln(t-t_0)-b)^2/2 a^2)}{(t-t_0)a{\sqrt{\pi}}}.
\end{equation}
This satisfies our boundary condition for $a = 0.5$. The scale parameter here is taken to $b= 1.5$
and $b=3.0$. 

\begin{figure}
\begin{center}
\includegraphics[angle=270,width=80mm]{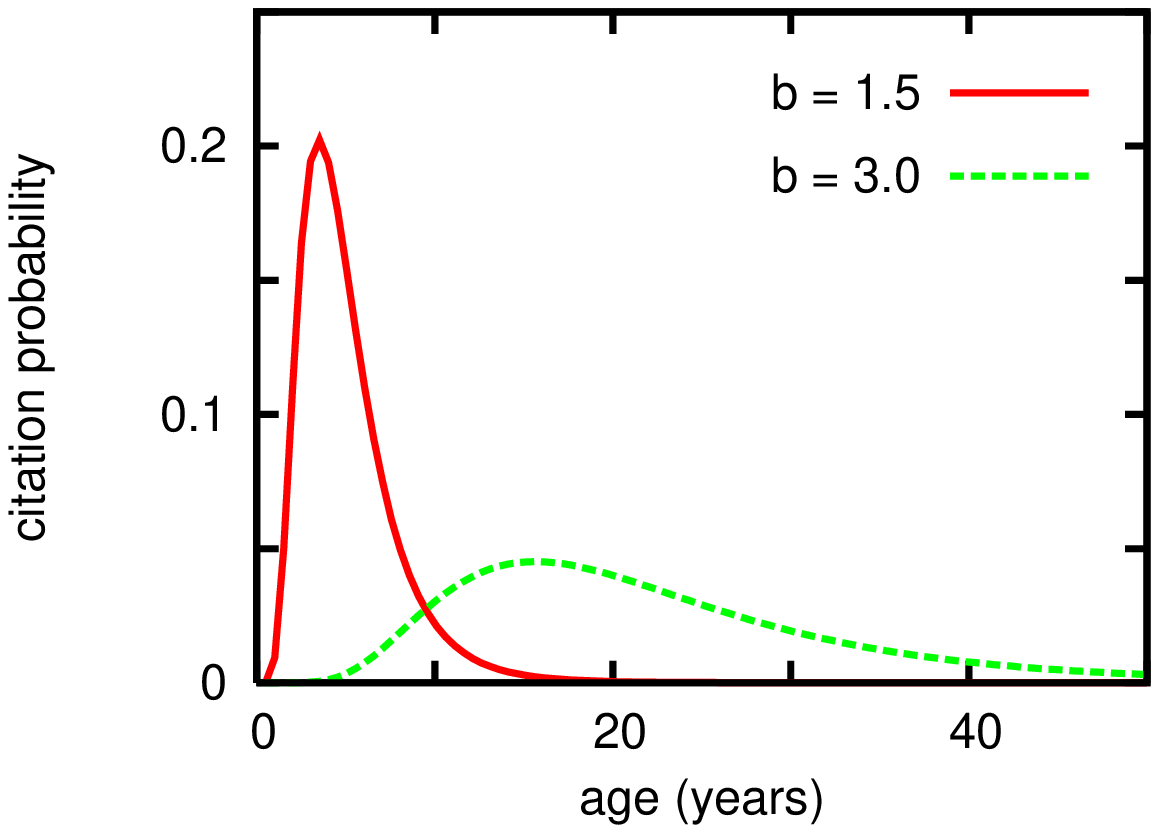}
\end{center}
\vskip -0.5cm
\caption{}{Citation lifetime distribution of an article for the lognormal distribution. 
The solid (red) line is for $b=1.5$ and the dashed (green) line is for $b=3.0$.}
\label{Fig.3}
\end{figure}
Fig. \ref{Fig.3} gives the citation lifetime distribution of an article for the lognormal distribution.
The x-axis is the age of the article and the y-axis gives it's probability of citation. 
The solid (red) line is for scale parameter $b=1.5$ and the dashed (green) line is for $b=3.0$.
Again, $b=3.0$ means that the discipline tends to cite older articles while $b = 1.5$ indicates that 
recent articles are more likely to be cited. The distributions are similar with some minor differences.
The rising of the curves are different. So the number of citations that articles accumulate in the 
initial years after publication determine whether the citation lifetime can be described by a Weibull 
or a lognormal distribution.

\begin{figure}
\begin{center}
\includegraphics[angle=270,width=80mm]{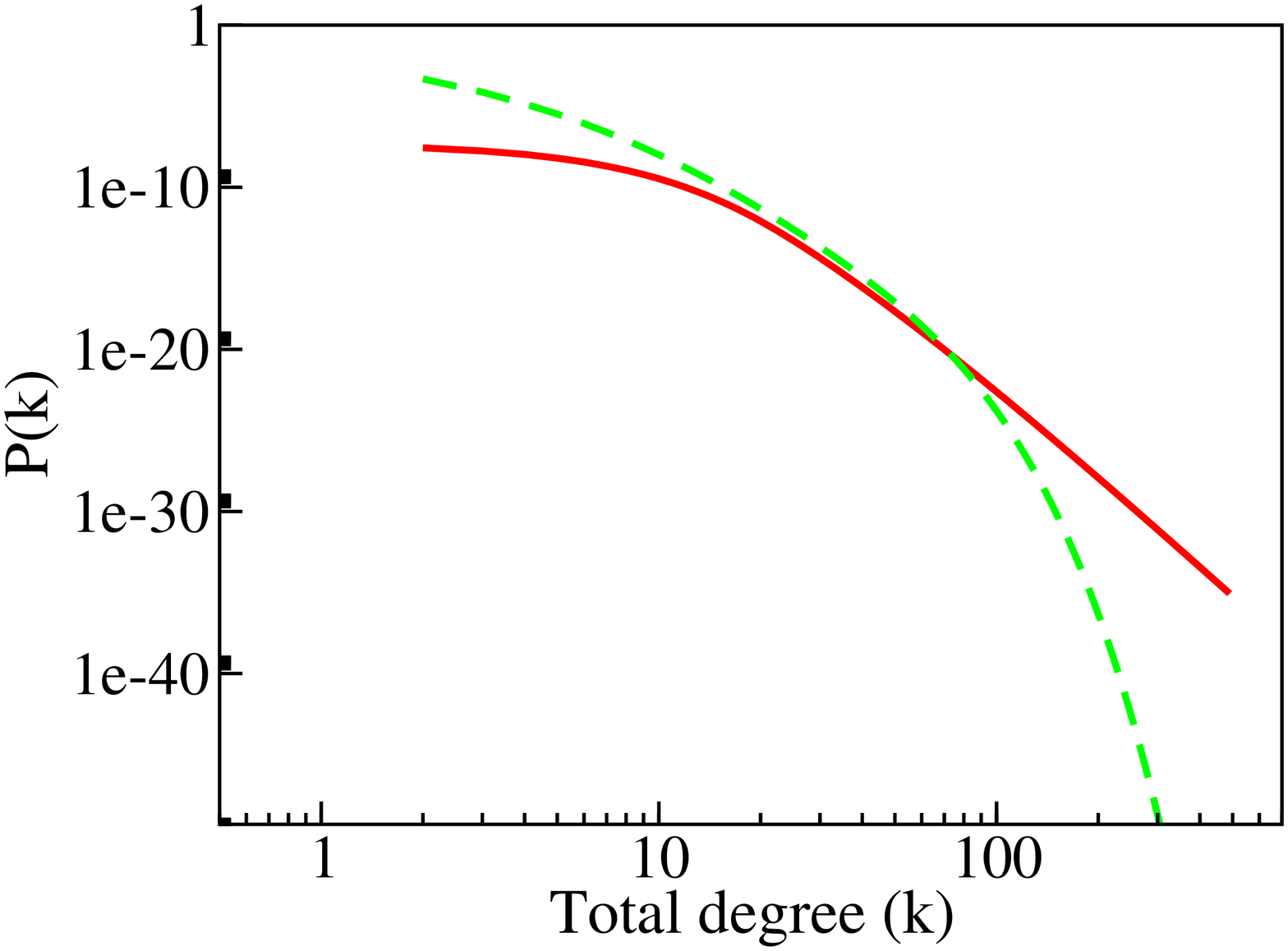}
\end{center}
\vskip -0.5cm
\caption{}{Degree distribution of citations to an article for different values of the scale parameter
for the lognormal distribution. The solid (red) line is for $b=1.5$ and the dashed (green) line 
is for $b=3.0$.}
\label{Fig.4}
\end{figure}
Fig. \ref{Fig.4} gives the total degree distribution for the lognormal distribution. 
As expected, it is similar to the one obtained for the Weibull. However, unlike the Weibull, here 
there are differences both in the initial  and the tail part for the different scale parameters.
The Weibull distribution with $b = 20$ and the lognormal distribution with $b = 3.0$ gives similar 
degree distributions. It is when $b$ is small, we see that the total degree distribution 
changes depending on the kind of lifetime distribution considered. This implies that if the average 
citation age of an article in a discipline is small, the number of citations received in the initial 
years after publication are important to determine the structure of the citation network in that 
discipline.

We have thus verified that citing patterns do indeed affect the final degree distribution of the 
network. The two lifetime distributions that we have chosen reflect the citation patterns in different 
disciplines. The two distributions differ in the rise of the citation probability with age. Though 
the scale parameter stretches both the distributions, for the lognormal distribution, the number of 
years before an article actually gets cited also changes. This changes the total number of citations 
accumulated in the initial years. The parameters of either of these distributions can be chosen to 
obtain a heavy tailed distribution. The scale parameter in that case would be small.  
Since the structure of the citation network depends on the parameters of the distribution, we conclude  
that the average age of citations in a discipline, the number of citations accumulated in the recent 
years and the tendency to cite more recent articles affect the topology of the citation network.

\subsection{Number of articles and references}
\label{sec:3}

Next we consider the average number of articles published each year and study how the increase 
in publications can affect the total degree distribution of the network. Increasing the number
of articles and increasing the time are similar since the average number of articles published 
each year in a discipline is considered constant. We find that increasing $m$ does not change 
the topology of the network. It only changes the numerical values obtained in the final degree
distribution.  The number of references cited by each article on the other hand does affect 
the topological structure of the network. This 
is particularly relevant as the number of references in different disciplines may vary considerably. 
The citation distribution is taken to be a Weibull distribution with the shape parameter $2$ and the 
scale parameter $b = 5$. The number of references $k_0$ considered are $20, 40,80,$ and $120$. 
Though review articles can have a much larger number of references, we consider the maximum value 
of $120$ since the number of articles generated per year in our case is kept constant at $200$.    
\begin{figure}
\begin{center}
\includegraphics[angle=270,width=80mm]{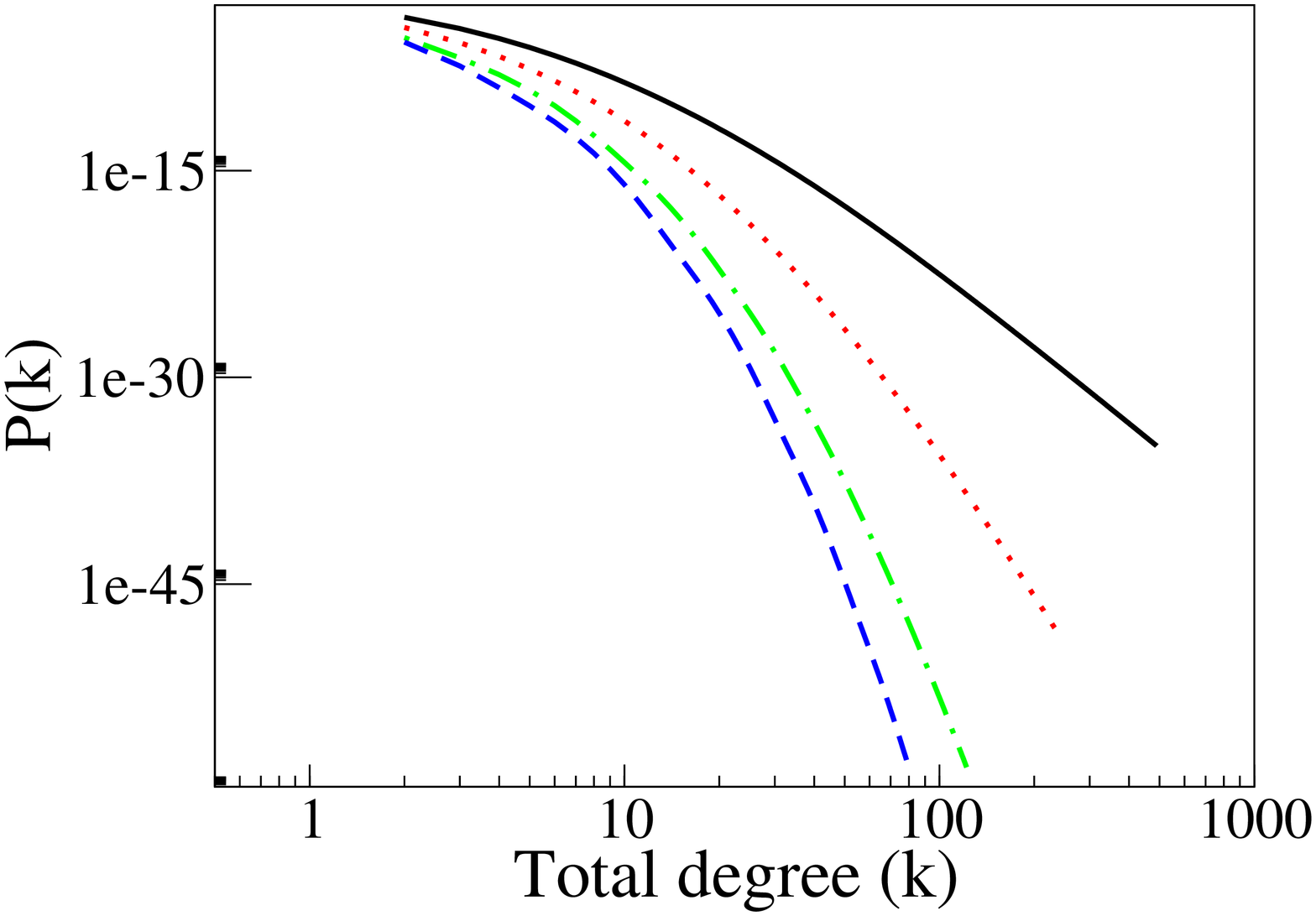}
\end{center}
\vskip -0.5cm
\caption{}{Total degree distribution for a Weibull distribution with $a = 2$ and $b = 5$ 
when the number of references vary. The straight (black) line corresponds to $k_0 = 20$, the dotted (red)
line to $k_0 = 40$, the dot-dash (green) line to $k_0 = 80$ and the dashed (blue)
line to  $k_0 = 120$. }
\label{Fig.5}
\end{figure}

Fig. \ref{Fig.5} gives the total degree distribution for different values of $k_0$. We find that a 
change in the number of references affects the network structure. For smaller number of references
the degree distribution can be described by a stretched exponential. As the average number of 
references increase, the distribution becomes exponential in nature (dashed curve in Fig. \ref{Fig.5}). Similar results are obtained for $b = 20$ and for the lognormal distribution. These indicate that the citation networks of  
two disciplines which have different citation practices will 
have different topological structure. The discipline which has a larger number of references per 
article will have deviations from the scale free structure and the total degree distribution will be 
given by an exponential instead of a power law.

\section{Real-world citation distributions}
\label{sec:IV}

We now discuss how our model relates to the various results obtained previously from different data sets. For a finite 
time period, we have identified three major parameters which determine the citation pattern in a 
discipline. These are (i) the number of 
references per article (ii) the average age at which articles get their maximum citations
and (iii) the rate at which the article accumulates citations in the initial years after
publication. This rate is given by the rise in the citation lifetime curve and determines
which distribution would be a better fit to the discipline. 
The study of the citation network of the Scientometric journal \cite{chun} over a 26
year time period generates two power laws with different scaling coefficients. As the authors 
point out, the number of articles published each year is relatively constant over this period.
So the deviations here are mostly due to the difference in the number of references per article.
They also find that the journal had a large number of self-referencing
in the first year. This went down steeply immediately after the first year and then increased 
steadily over time in the next 25 years. 
According to our model, as the number of references per article increases, the slope of the 
degree distribution on a logarithmic scale becomes steeper. So the change in the power law 
coefficient in the degree distribution could be due to the increase in the number of references
within the journal. Results obtained from the SPIRES database \cite{lehmann} also show that 
the degree distribution in high energy physics can be fitted by two power laws independent of each other. 
As mentioned previously, the total degree distribution obtained from the Weibull distribution 
with shape parameter $a = 2$ and scale parameter $b = 5$ can also be fitted by two power laws independent of 
each other. 

These values of the parameters are not unique, other combinations also generate similar 
curves but with different scaling coefficients.  So it may well 
be that the citation pattern in each of these datasets are fit with a Weibull or lognormal distribution 
with different parameters. Our motivation for using lifetime distributions to illustrate the 
importance of citation pattern on network evolution is largely theoretical. For a better validation 
of the model it is necessary to obtain the distribution from an actual dataset and fit a probability 
distribution. The parameters obtained from the fit can then be used to predict the degree distribution 
of the final network. We do this for citations obtained from the Physical Review D journal.

\begin{figure}
\begin{center}
\includegraphics[angle=270,width=80mm]{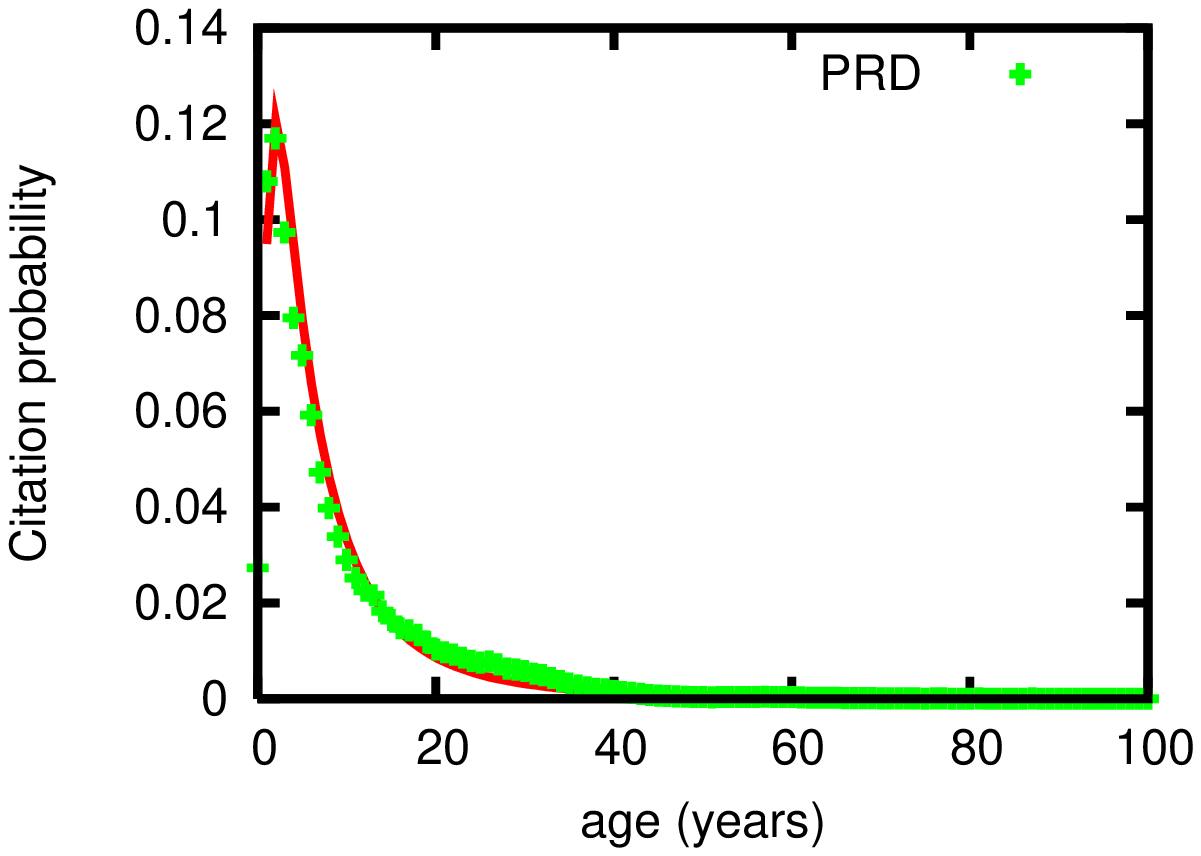}
\end{center}
\vskip -0.5cm
\caption{}{The citation pattern in the Physical Review D journal is fitted with a lognormal 
distribution with $a = 1.1$ and $b=1.7$ }
\label{Fig.6}
\end{figure}

The degree distribution of the citation network for the Physical Review D journal has previously 
been analyzed extensively \cite{redner1,redner2,vazquez}. However the citation pattern has not been 
obtained before. We obtain the  citation pattern by plotting the number of citations 
received by an article against the age of the article. This data is then normalized and fitted with a 
probability distribution. Fig. \ref{Fig.6} gives the plot. The probability distribution turns out to 
be a lognormal distribution with the parameters $a =1.1$ and $b = 1.7$.

\begin{figure}
\begin{center}
\includegraphics[angle=270,width=80mm]{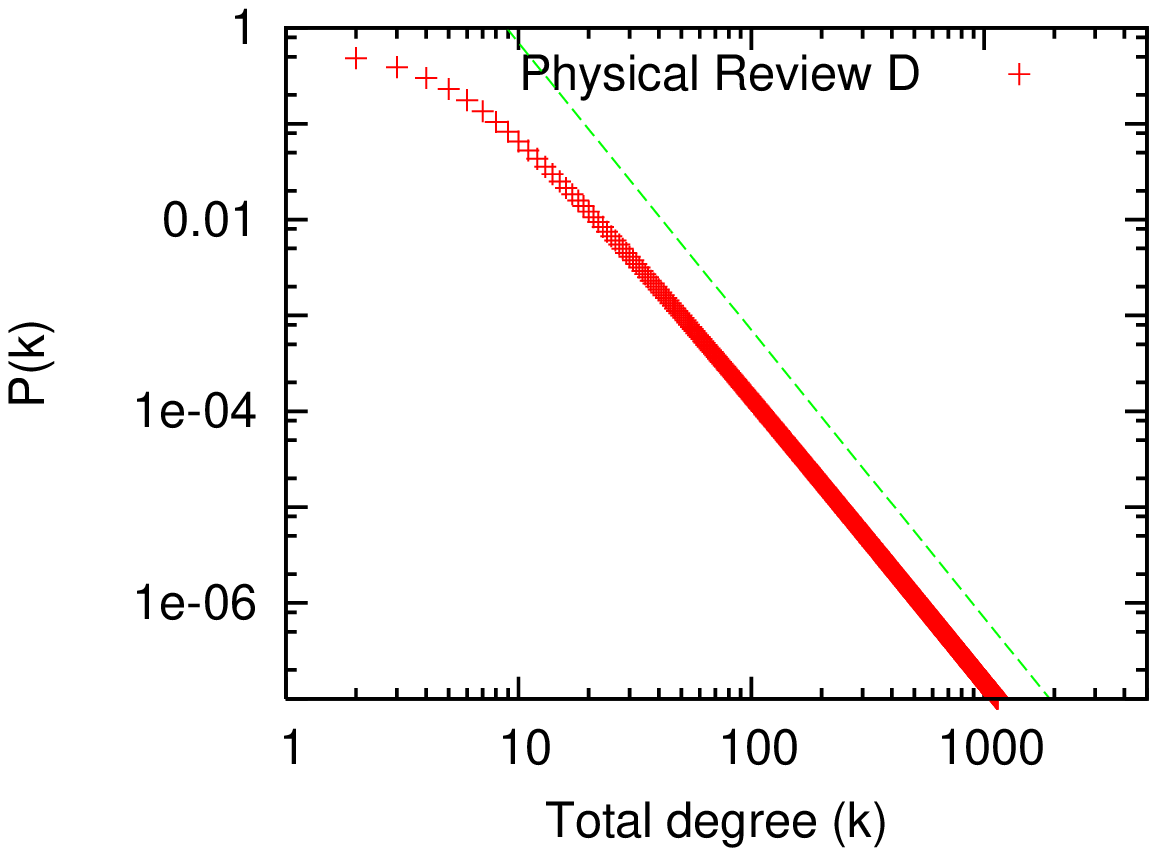}
\end{center}
\vskip -0.5cm
\caption{}{Degree distribution obtained from Eq. \ref{eq:soln} for Physical Review D. 
The number of articles per year is $1000$ and the average number of references is $4$. The line denoting the power law 
is shifted upwards for better visualization of the data and the fit.}
\label{Fig.7}
\end{figure}

Once $f(t-t_0)$ is obtained, we calculate
the citation probability from  Eq. \ref{eq:soln}. 
We plot this in Fig. \ref{Fig.7}. Since we are considering references in Physical Review D journals 
only, we take the average number of articles published per year to be $1000$. 
The Physical Review D started in 1970. In the initial years 
of publication, the average number of articles was very low (about $300$) but in recent years the 
number has increased to around $2000$. Since our average number of articles generated each year 
remains constant throughout the time period considered (1970 - 2004), we take $m = 1000$. 
To obtain a scaling co-efficient of $-3$ in the tail region (as obtained in Ref.\cite{redner1}) with 
$m = 1000$, we find that the number of references should be $4$. The total degree distribution 
and the power law with an exponent of -3.0 is shown in Fig. \ref{Fig.7}. 

Since an average number of references of $4$ seemed quite small, we went back to the dataset
and examined the references in the journal Physical Review D. A random sampling indicated that the 
initial articles in the journal have few or no citations. Later articles have an average of $20$ 
references per article but most of the articles cited are published in other journals. The average 
number of references to Physical review D is less than $10$. The citation lifetime was obtained 
from the references from articles published in the Physical Review D irrespective of where the 
{\it cited} article was published. However, the degree distribution in Ref.\cite{redner1} did 
not include citations to articles outside the Physical Review journals. So the average number of references 
does not seem anomalous. It is the average number of references to Physical Review journals in 
Physical Review D. Since many of the initial articles in the journal had references to journals 
cited outside the Physical Review, the average number of references in the total time period is 
reduced.

In our model we have considered the number of articles published and the average number of references
per article to be constant throughout the time period. This is not true for real 
world citation data. We have shown that changing the number of references 
changes the scaling coefficient, since in the real world data set these number keep changing, it becomes
difficult to fit any citation data with an unique power law.  
This is further illustrated in Lehmann's work \cite{lehmann}, where he plots the degree distribution for the 
total SLAC SPIRES database and also plots the degree distribution separately for the subfields 
in the database. The citation pattern in the subfields being more homogeneous, the degree distribution 
here can be fit by two power laws whereas the degree distribution for the entire database can be fit 
by multiple power laws.

\section{Summary and conclusions}
\label{sec:V}

We have presented an analytical model for an evolving citation network that takes into account the 
citation pattern in a particular discipline. Though earlier models have considered the aging of 
articles and preferential attachment, our model is the first to include 
definite characteristics that define the citation pattern in a discipline. Apart from the 
citation pattern, we also include the number of articles published per year. 
We find that all these together determine the total degree distribution of the network.

We propose modeling the aging probability by a lifetime distribution. Two of the most commonly 
used lifetime distributions are the Weibull and the lognormal.
The Weibull distribution has been used previously in the TARL model \cite{borner}. It was used to 
model the aging characteristics of the citation network obtained from the PNAS dataset. We also 
found that the aging characteristics of the citation network in the Physical Review dataset fits
a lognormal distribution. The choice of lifetime distributions to model the aging probability is thus 
consistent with real datasets. 
 
We find that the topological structure of the network depends on the citation practices followed in 
different 
disciplines. Citation practices vary considerably between different disciplines. The average age of the 
cited article and the number of references per article seem to be the two most important factors which 
determine the scaling coefficient of the power law. We find that the deviations from the power law 
structure are due to aging and the number of references per article. We use the Physical Review D 
dataset to validate our model. We consider an average of $1000$ articles published in a year. 
A power law with exponent $-3$ is obtained for an average of $4$ references per article. 
The low number of references is not altogether anomalous because it refers to the average 
number of reference to Physical Review journals only. However, there are many assumptions in the 
model which may have lead to this value. We now discuss these assumptions and discuss their 
impact on our model.

The probability of citation based on preferential attachment and aging are considered to be independent 
of each other. Though this is an accepted assumption in modeling citation networks, it is not 
true in the real world \cite{redner2}. Highly cited articles will definitely have a longer lifetime 
compared to the less cited articles in any discipline. In Ref. \cite{redner2} the average citation 
age for most Physical Review papers is found to be $6.2$ years, while the average citation age for
highly cited papers is found to be $11.7$ years. This means that the two probabilities should be 
dependent on each other, even if the dependency is small. Since the total degree distribution 
of our model depends on the citation pattern, a change in the citation pattern will definitely 
change the final structure of the network.  

We have also assumed that the number of 
articles published each year  and  the number of references per article are constant. 
For real world citation data neither is true. In fact in some cases the increase in the number of 
articles published each year is exponential. This is especially true for the Physical Review D journal 
which we have used to validate our model. As mentioned previously, the number of articles 
published each year in this journal has grown by an order of magnitude in three decades. This 
unfortunately is not reflected in our model which considers an average value for the whole 
time period.  However, our point was to exemplify the importance of citation patterns in the 
evolution of a citation network. It is encouraging to find that our final degree distribution is 
similar to the degree distribution obtained in ref.\cite{redner1}. While our model is able to explain 
the differences observed in different citation networks qualitatively, we hope to improve it further 
and make quantitative predictions about citation patterns observed in different disciplines.

{\bf Acknowledgments}  
This work is supported by the James S. McDonnell Foundation grant in the area 
Studying Complex Systems entitled "Modeling the Structure and Evolution of Scholarly Knowledge"
and the Cyberinfrastructure for Network 
Science Center. It has greatly benefited from discussions with Katy B\"{o}rner, and Robert Goldstone and critical comments from Debasis Dan and Shashikant Penumarthy. 
The author thanks Shashikant Penumarthy for providing the data used for obtaining the citation pattern 
for the Physical Review D journal.

\end{document}